\documentclass[runningheads]{llncs}
\usepackage{graphicx}

\usepackage{tikz}
\usepackage{comment}
\usepackage{amsmath,amssymb,bbm} 
\usepackage{color}
\usepackage{times}
\usepackage{epsfig}
\usepackage{graphicx}
\usepackage{amsmath}
\usepackage{xcolor, url}

\DeclareMathOperator*{\argmax}{argmax}
\DeclareMathOperator*{\argmin}{argmin}

\newcommand{\subhead}[1]{ 
	\vspace{0.2em}
	\noindent \textbf{{\smash{#1}}}:  
	}

\newcommand{\method}[1]{ 
	\underline{\smash{#1}}: }

\newcommand{\boldhead}[1]{\textbf{#1}: }

\begin{document}
\pagestyle{headings}
\mainmatter

\title{Defenses Against Multi-Sticker Physical Domain Attacks on Classifiers} 

\titlerunning{Defenses Against Multi-Sticker Attacks}
%
\author{Xinwei Zhao \orcidID{0000-0002-4328-4846} \and
	Matthew C. Stamm \orcidID{0000-0002-3986-4039}}
\authorrunning{X. Zhao \and M C. Stamm}
%
\institute{Drexel University, Philadelphia, PA, USA \\
	\email{xz355@drexel.edu, mstamm@drexel.edu}}
\maketitle

\begin{abstract}
Recently, physical domain adversarial attacks have drawn significant attention from the machine learning community. One important attack proposed by Eykholt et al. can fool a classifier by placing black and white stickers on an object such as a road sign. While this attack may pose a significant threat to visual classifiers, there are currently no defenses designed to protect against this attack.  In this paper, we propose new defenses that can protect against multi-sticker attacks.  
We present defensive strategies capable of operating when the defender has full, partial, and no prior information about the attack. By conducting extensive experiments,  we show that our proposed defenses can outperform existing defenses against physical attacks when presented with a multi-sticker attack.

\end{abstract}
\section{Introduction}

Deep neural networks have been widely used for many visual classification systems, such as autonomous vehicles~\cite{geiger2012we,urmson2008autonomous} and robots~\cite{zhang2015towards}.However,  deep neural networks are vulnerable to adversarial attacks~\cite{carlini2017towards,chen2018ead,2014arXiv1412.6572G,karmon2018lavan,kos2018adversarial,liu2016delving,madry2017towards,moosavi2017universal,DeepFool_2016_CVPR,nguyen2015deep,papernot2016limitations,su2019one,szegedy2013intriguing}. By modifying the pixel values of an image,  many classifiers can be fooled.

Recently, attacks that can operate in the physical world have started to attract increasing attention~\cite{athalye2017synthesizing,brown2017adversarial,eykholt2018robust}. While some physical domain attacks require  crafting a new object~\cite{athalye2017synthesizing,eykholt2018robust,kurakin2016adversarial}, other attacks can fool the classifiers by adding one or a few physical perturbations, such as printable patches~\cite{brown2017adversarial,eykholt2018robust} on or next to an object.  The adversarial patch attack creates one universal patch that can be used to attack an arbitrary object once it is trained, regardless of scale, location and orientation~\cite{brown2017adversarial}.  
The camouflage art attack uses black and white stickers that are applied to an object such as a traffic sign to make a classifier believe it is a different object.
~\cite{eykholt2018robust}
Since these physical perturbations are very concentrated and confined to small regions,  it is easy for  attackers to craft these physical perturbations and put the attack in practice in the real world. 

Previous research shows that defenses against digital domain attacks~\cite{bastani2016measuring,bhagoji2017dimensionality,chiang2020certified,das2018shield,dhillon2018stochastic,dziugaite2016study,feinman2017detecting,guo2017countering,hayes2018visible,levine2020randomized,madry2017towards,naseer2019local,papernot2016distillation,raghunathan2018certified,shaham2018defending,xu2017feature} may not be able to defend against physical domain attacks, such as the camouflage art attack,  because physical perturbations are usually stronger than those produced by digital domain attacks. 
Some recent research  has been done to defend against physical domain attacks~\cite{chiang2020certified,hayes2018visible,levine2020randomized,naseer2019local,xu2019lance}.  
\begin{figure}
	\centering
	\includegraphics[scale=0.4]{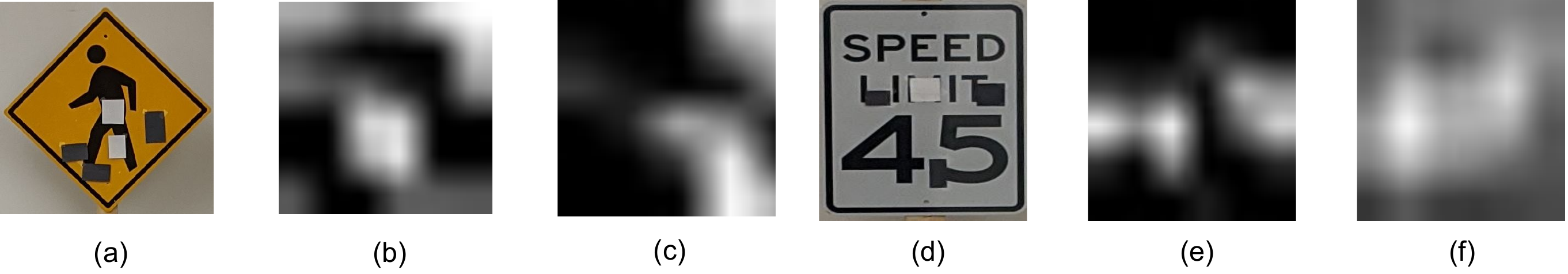}
	\caption{Attacked signs (a) \& (d) as well as their Grad-CAM activation maps before  attack (b) \& (e) and after  attack (c) \&~(f).}
	\label{fig: gradcam}
\end{figure} 

Existing research, however,  focuses on defending against adversarial patches, and  does not translate to defend against other physical attacks like the camouflage art attack (i.e white and black sticker attack).
For example, one  approach to defend against the adversarial patch  attack is to first locate the  perturbed area using an attention-based or gradient-based model, and then remove or diminish these areas~\cite{hayes2018visible,naseer2019local,xu2019lance}. 
The perturbations produced by multi-sticker attacks like the camouflage art attack, however, cannot be detected the same way due to several reasons.
First,  the black and white stickers  produced camouflage art attack are not highly textured, and hence are unlikely to  be detected via gradient-based methods. 
Second, the camouflage art attack works in conjunction with the scene content to  redirect the classifiers decision instead of hijacking its attention like the adversarial patch does.  As a result, multi-sticker attacks are unlikely to be identified using attention-base models. 

An example of this phenomenon can be seen in Figure~\ref{fig: gradcam}, which shows activation maps produced by Grad-CAM~\cite{selvaraju2017grad}  when presented with images before and after a multi-sticker attack.
When examining the activation maps of the pedestrian crossing sign before an attack shown in Figure~\ref{fig: gradcam}(b) and after the attack shown in Figure~\ref{fig: gradcam}(c), we can see that the attack has shifted the classifier's attention off of attacked sign.  Defenses that operate by removing or altering these regions will have no effect on the attack.  
Alternatively, from examining the activation maps of an unattacked speed limit sign in Figure~\ref{fig: gradcam}(e) and it's attacked counterpart in Figure~\ref{fig: gradcam}(f), the classifier is paying attention to nearly the entire sign.  Defenses that operate by removing or distorting these regions will degrade the image so severely that the classifier will be unable to operate.

Furthermore, it is important for defenses against physical domain attacks to be evaluated on real images of physically attacked objects.  Digital simulations of physical attacks are sometimes used for evaluation due to the ease of creating a dataset, for example, digitally adding perturbations that simulate a physical attack into an image.  However, these digital simulations do not capture many effects that occur during imaging, such as lighting conditions, the curvature of surfaces, focus blur, sampling effects,  etc.   In practice, phenomena such as these can impact how a camera captures physical domain perturbations, and can potentially affect the success of defenses.  Defenses that are highly tuned to features of “pristine” digital simulations of attacks may be less successful when confronted with real images of physically attacked objects or scenes. 

In this paper, we propose a new defense strategy that does not rely on attention models to identify attacked image regions and can successfully defend against multi-sticker attacks, like the camouflage art attack. Our proposed defense operates by first creating defensive masks that can maximize the likelihood of guessing the location of the perturbations, then mitigates the effect of the perturbations through targeted modifications, and eventually make a final decision based on  defended images. 

\noindent
\textbf{Our Contributions}: 
\begin{itemize}
\item We propose a set of new defenses that can protect against multi-sticker physical domain attacks such as the camouflage art attack by Ekyholt et al.~\cite{eykholt2018robust}.  To the best of our knowledge, no existing defenses are designed to defend against such attacks.

\item We present practical defenses that can be utilized depending on whether the defender has full knowledge of the attack (non-blind), partial information about the attack (semi-blind), or no information regarding the attack (blind).

\item We create a new database of front-facing photos of  90 physically attacked signs using camouflage art attack and use this database to assess our defense. 

\item We demonstrate that our proposed defenses outperform other state-of-the-art defenses against physical attacks, such as the digital watermark defense ~\cite{hayes2018visible}, when presented with multi-sticker attacks.
\end{itemize}

\section{Additive Physical Domain Attacks}
Adversarial attacks  pose an important threat against deep neural networks~\cite{athalye2017synthesizing,brown2017adversarial,carlini2017towards,chen2018ead,eykholt2018robust,2014arXiv1412.6572G,karmon2018lavan,kos2018adversarial,kurakin2016adversarial,liu2016delving,madry2017towards,moosavi2017universal,DeepFool_2016_CVPR,nguyen2015deep,papernot2016limitations,su2019one,szegedy2013intriguing}. Some physical domain attacks, like the adversarial patch~\cite{brown2017adversarial} and the camouflage art attack~\cite{eykholt2018robust},  have shown that  adding perceptible but localized patches to an object can make a classifier identify it as a different object.  We now briefly describe how these two physical domain attacks are launched at a classifier $C$ using attack target class $t'$.

\subhead{Adversarial patch}
To generate an adversarial patch $A'$, the authors of~\cite{brown2017adversarial} use an operator $O(I, A, \theta_l,\theta_t)$ to transform a given patch $A$, then apply it to an image $I$ at location $\theta_l$.  Similarly to an Expectation over Transformation attack (EoT)~\cite{athalye2017synthesizing}, the adversarial patch can be obtained by optimizing over sampled transformation and locations, 
\begin{equation}
A'=\max_A \mathbb{E}_{I\sim \mathcal{I}, \theta_l\sim\Theta_L, \theta_t\sim\Theta_T} C(t'|O(I, A, \theta_l, \theta_t)) 
\end{equation}
where $\mathcal{I}$ denotes the training image dataset,  $\Theta_T$ denotes the distribution of transformation and $\Theta_L$ denotes the distribution of the location. 
Once the patch is trained, it can universally attack any object.

\subhead{Camouflage art attack} 
Launching the camouflage art attack involves finding a single set of perturbations $P$ that are capable of fooling
a classifier under different physical conditions.  
This attack, which produces perturbations for a given pairing of source and target class, 
was demonstrated by using it to fool a classifier trained to distinguish between different US traffic signs.
Let $H^v$ denote the distribution of the image of an object under both digital and physical transformations,  and $h_i$ denote  each sample from this distribution. The attack perturbations can be obtained via optimizing, 
\begin{equation}
\argmin_P\lambda||M_h, P||_p + \mathbb{E}_{h_i\sim H^v}J(C(h_i+G(M_h, P), t')
\end{equation}
where  $M_h$ is the mask that applies spatial constraints to the perturbation (i.e ensures the perturbation is within the surface area of the object), $\lambda$ is a hyper-parameter that regularize the distortion,  $J(\cdot)$ is the loss function that measures the difference between the classifier's prediction of the attacked object and the target class, $G(\cdot)$ is the alignment function that  maps transformations on the object to transformations on the perturbation, $||\cdot||_p$ denotes  $\ell_p$ norm.
\section{Problem Formulation}
%

We assume that the system under attack wishes to analyze some scene $S(x,y)$ containing an object to be classified.  To do this, the system will capture a digital image $I(x,y)$ of the scene, which will then be provided to a pre-trained classifier $C(\cdot)$ which maps the image into one of $N$ classes $t\in\mathcal{T}$.
For the purposes of this work, we assume that if no adversarial attack is launched, then the image provided to the classifier is $I=S$.

An attacker, may attempt to fool the classifier by launching a physical domain attack $\alpha(\cdot)$.  This corresponds to physically modifying an object within the scene by adding adversarial perturbations $P$ to it.
Since these perturbations must be physically added to the scene, we assume that they will be spatially localized to one or more regions of the object under attack.  These regions can be specified by a spatial mask $M$, 
where $M(x,y)=1$ corresponds to a perturbation being present at  spatial location $(x,y)$ and $M(x,y)=0$ corresponds to no perturbation occurring at $(x,y)$.
As a result, we can express a physically attacked scene  $\alpha(S)$
\begin{equation}
\alpha(S(x,y))= (1-M(x,y)) S(x,y) + M(x,y) P(x,y).
\label{eqn:physAttackModel}
\end{equation}

In this paper, we assume that the adversarial perturbations will take the form of black and white stickers added to an object as proposed by Eykholt et al.~\cite{eykholt2018robust}, i.e. $P(x,y)=\{black,white\}$.
  Other physical domain attacks, such as the adversarial patch~\cite{brown2017adversarial} 
 can still be modeled using \eqref{eqn:physAttackModel} by allowing $P(x,y)$ to correspond to the full range of color values.  Since the majority of the defenses proposed in this paper do not rely on knowledge of the color values of $P$, it is likely that these defenses can be used against other physical domain attacks such as the adversarial patch.  
We note that this work only addresses physical domain attacks that involve modifying an existing physical object, and not attacks that involve the creation of a new physical object such as synthesized 3D objects~\cite{athalye2017synthesizing} and printed photos or posters~\cite{eykholt2018robust,kurakin2016adversarial}.

\section{Knowledge Scenarios}
\label{sec: scenarios}
To defend a classifier, we first assume that the defender has full access to the classifier and implicitly knows the $N$ classes that it is trained to distinguish between. 
We examine three scenarios corresponding to different levels of knowledge available to the defender.

\subhead{Non-blind}
We assume that defender knows if an object is attacked or not,  the perturbation masks $M$ that indicates the perturbation areas and the perturbations $P$.  Therefore, locations of  perturbations can be directly located.

\subhead{Semi-blind}
We assume that the defender does not know if the object is attacked or not. We also assume that if the object was attacked, the defender does not know the perturbation masks $M$. However, the defender knows the attack method $\alpha(\cdot)$. Therefore, for any source A and target B pairing,  the defender can obtain a perturbation mask $M_{\text{A, B}}$ via launching the attack.

\subhead{Blind}
We assume that defender has zero knowledge. Specifically, the defender does not know whether an object is attacked or not.  We also assume that if the object was attacked, the defender does not know the perturbation regions. Additionally, the defender does not know the attack method.

\section{Proposed Defenses}
\label{sec: proposed}
To defend against a physical domain attack,  we  propose a set of defenses based on the amount of knowledge available to the defender. These defenses attempt to interfere with or remove adversarial multi-sticker perturbations to mitigate their effects.  If the defender is able to leverage information about the potential locations of these perturbations, defenses are guided to these regions.  Otherwise, our defenses are designed with the intuition that adversarial perturbations are more fragile to distortions than the underlying object that they are attacking.

Our defensive strategy is composed of three major steps.  First, we obtain a defensive mask $R$ or set of defensive masks $\mathcal{R}$ indicating candidate areas to apply defenses.  Second, we launch a local defense in regions indicated by a defensive mask to produce a defended image $\delta$.  When our first step results in a set of defensive masks, local defenses can either be sequentially applied in conjunction with each mask to produce a single defended image, or they can be applied in parallel to produce a set of defended images.  In the third step, the defended image or images are provided to the classifier.  If a set of defended images are produced by the second step, a fusion strategy is employed to produce a single classification decision. In what follows, we discuss each step of our proposed defenses in detail.

\subsection{Defensive Mask Selection}
The goal of each defensive mask is to ensure that defensive distortions are only applied to small regions of the image, since each perturbation produced  by the multi-sticker attack is still confined to a small region. We do not want to change the ground truth object.  Let $R(x,y)\in \{0,1\}$ denote a defensive mask, where 1 indicates the area need to be defended, 0 indicates the area of the ground truth content. Now we discuss the acquisition of defensive masks. 

\subhead{Oracle Approach} 
If the perturbation mask $M$ is known, such as in the non-blind scenario,  we simply let $R=M$.

\subhead{Estimated Defensive Mask Sets} 
In semi-blind scenarios, the defender may know the potential attack method $\alpha$, but not perturbation masks or the potential attack mask if the attack was launched. They can, however, leverage knowledge of $\alpha$ to create a set of estimated defensive masks.

To do this, first we assume that $I$ is an image of an attacked scene.  The attack's target class $\hat{t}$ can be inferred by using $C$ to classify the image 
such that $\hat{t}=C(I)$.  
Next, the defender can create their own implementation of $\alpha$ and use it to recreate an attack aimed to move true class $j$ to target class $\hat{t}$.  The attack's perturbation mask can then be used as the estimated defensive mask $R_{j,\hat{t}}$ for source $j$ and target $t$.  This process can be repeated for all $j \in \mathcal{T}$ such that $j \neq \hat{t}$ to produce the set of estimated masks ${\mathcal{R}_{\hat{t}}=\{R_{1, \hat{t}}, \ldots, R_{\hat{t}-1,\hat{t}}, R_{\hat{t}+1,\hat{t}}, \ldots, R_{N, \hat{t}}\}}$. 
To reduce computational costs while launching the defense, the set $\mathcal{R}_{\hat{t}}$ can be precomputed for each target class. With increasing number of classes, the computational cost may become high  for constructing sets of estimated set of defense masks  and launching the defense. To solve this problem, defender can use a subset of defensive masks instead of every single mask. We propose two methods to form these subsets.

\underline{\smash{Ranked Selection}}:
The defender can utilize class activations to guide the selection of the subset of defensive masks to use.
Since physical attacks operate by constraining  perturbations to small areas to avoid suspicion, it is reasonable to assume these perturbation push the object just across the boundary of its true class. Therefore, the true class of an attacked image most likely shows up in the top few activated classes. 
To guide the selection of defensive masks, first we assume that a scene is always under attack (regardless of whether this is true or not) 
and treat the class with the highest activation as the target class.
The true class then lies among the remaining classes,  which are ranked according to their activation scores.
The subset of $k$ defensive masks  is then chosen as the set of masks created using the assumed target class (i.e. the class with the highest activation) and the $k$ top candidates for the true source class (i.e. the classes with the second highest through $k+1$ highest activations).  
By doing this, the defender can control the computation cost of the defense while increasing the chance that the most useful defensive masks are utilized.

\underline{\smash{Random Selection}}:
A heuristic way to form a subset of  defensive masks is through random selection. Since each of the selected mask is related to the target class, each selected mask can be used to defend a partial of the image. By grouping several defensive masks, it  may increase the chance for a  successful defense.

\subhead{Randomly Chosen Regions} In blind scenarios, the defender cannot leverage any prior information about the attack or possible perturbation locations.  
In these situations, we create a set of defensive masks made by randomly choosing defensive regions.  Our intuition is that if we use many random defensive masks, several of them will interfere with the adversarial perturbations. 
Each mask is made by randomly selecting $m$ different $w\times w$ windows to apply localized defenses.  We use two different approaches for randomly choosing these regions:

\underline{\smash{Overlapping}}:  
The locations of each window are chosen uniformly at random from throughout the image area.  As a result, some windows may overlap with one another.

\underline{\smash{Non-overlapping}}:
In this approach, we ensure that defensive regions are spread throughout the region by disallowing overlaps.  We do this by first dividing the defensive mask into non-overlapping $w\times w$ blocks, then randomly choosing $m$  of these blocks as defensive regions.

\subsection{Local Defense Strategies}
\label{subsec: LDS}
After the defensive masks are obtained, we can apply local defenses to image regions specified by these masks.  
To make it clear,  we first show how to obtain the defended image using single defensive mask, then we adapt the proposed defenses to accommodate multiple defensive masks.

Given one defensive mask, we propose two methods to defend against the attack. 

\subhead{Targeted perturbation remapping} This idea is to interfere the perturbations instead of removing it.  Specifically, we can using remapping functions  to destroy the spatial correlation between perturbed regions. Let $\phi(\cdot)$ be the remapping function, then a single defended image can be expressed as,  
\begin{equation}
\delta(x, y)=\begin{cases} 
I(x, y) & R(x, y)=0 \\
\phi(I(x, y)) & R(x,y)=1
\end{cases}
\end{equation}
 In this work, we consider three mapping functions:
 
 \method{RemapW} Change pixels to  white.

\method{RemapB}  Change pixels to black.
 
\method{RemapT} Pick a threshold $\tau$, change pixels to black if the luminance value is above the threshold and to white if below the threshold.

\subhead{Localized region reconstruction} The idea is to diminish or remove the effects of that perturbation by reconstructing perturbed local regions of input image on the basis of other parts of the image.  Since the perturbations are confined to a small region, we can use the inpainting algorithm to reconstruct the image.

The defenses discussed above can be easily adapted for multiple defensive masks. Let $\psi(\cdot)$ denote the defense. For a set of defensive masks $\mathcal{R}$ that comprises $k$ mask, $\mathcal{R}=\{R_1, R_2, ..., R_k\}$, we can either obtain one single final defended image via sequential defense, or obtain a sequence of individually defended images via parallel defense and then fuse the results.  Now we discuss sequential and parallel defense individually. 

\subhead{Sequential defense} We attempt to make the defense stronger by recursively applying the defense and obtain a single defended image.  For iteration $\ell$, the defense $\psi(\cdot)$ is applied to the output of the previous step using $\ell^{th}$ defensive mask, $\psi_\ell(\cdot)=\psi(\delta_{\ell-1}, R_\ell)$. The final defended image is obtained by  sequential applying the defense using each of $k$ individual defensive mask via, 
\begin{equation}
\delta=(\psi_k\circ \psi_{k-1} \circ \ldots \circ \psi_1)(I)
\end{equation}

\subhead{Parallel defense} The idea is to generate many copies of defended image with each copy being able to defend one part of input image. Using $\ell^{th}$ defensive mask,  we define $\ell^th$ defended image as $\delta_\ell=\psi(I, R_\ell)$, then using $k$ defensive masks we get $k$ individual defended images, $\{\delta_1, \delta_2, \ldots, \delta_k\}$. 

\subsection{Defensive Classification}
After applying local defenses,  we need to use the classifier to make a final decision on the defended image or images. We propose two decision making strategies. 

\subhead{Single defended image} After the sequential defense, the defender will obtain  a single defended image.  We simple use the classifier to classify the defended image, 
$t=C(\delta)$. 

\subhead{Multiple defended images} The parallel defense will result in a sequence of defended images. The defender can use the classifier to get a fused decision by combining the decisions of the individually defended images. We propose two fusion strategies.  

\method{Majority vote (MV)} Use the classifier to make a decision with each individual defended image, $t_\ell=C(\delta_\ell)$, then take a majority votes of all decisions
\begin{equation}
t=\argmax_{n\in N} \sum_{\ell=1}^k\mathbbm{1}(C(\delta_\ell)=t_n)
\end{equation}
where $\mathbbm{1}$ is the indicator function.

\method{Softmax fusion (SF)}  Let $\mathbf{v^{(\ell)}}$ denote the softmax output of the classifier for the $\ell^{th}$ defended image,  $\mathbf{v}^{(\ell)}=C_{softmax}(\delta_\ell)$, next add the softmax output of each of the $k$ defended images to form a single vector $\mathbf{v}$, 
\begin{equation}
\mathbf{v}=\sum_{\ell=1}^k\mathbf{v^{(\ell)}}
\end{equation}
 then take the class corresponding to  the largest value in $\mathbf{v}$ as the final decision,
\begin{equation}
t=\argmax_{n\in N} v_n
\end{equation}
where $v_n$ is the $n^{th}$ element in the vector $\mathbf{v}$.

\section{Evaluation Metrics}
When formulating our evaluation metrics, we let $t^*$ denote the ground truth class of a scene.  Additionally, we let $\pi_A$  denote the a priori probability that an attack is launched against a scene.

 \subhead{Classifier} To evaluate the baseline performance of the classifier $C(\cdot)$,  we calculate the classification accuracy as the probability that the image  of a scene being correctly classified as its ground true class,
 \vspace{-0.5em}
\begin{equation}
\text{CA}=Pr(C(I)=t^*|I=S)
\end{equation}
\subhead{Attack} To evaluate the baseline performance of the attack,  we calculate the targeted attack success rate (T-ASR) and  the untargeted attack success rate(U-ASR). 

T-ASR is defined as the probability that the image of an attacked scene  is classified as the target class,
\begin{equation}
\text{T-ASR}=Pr(C(I))=t'|I=\alpha(S))
\label{eq: t-asr}
\end{equation}  
\vspace{-0.5em}

U-ASR is defined as the probability that the image of an attacked scene  is classified as any other class than the true class,
\begin{equation}
\text{U-ASR}=Pr(C(I))\neq t^*|I=\alpha(S))
\label{eq: u-asr}
\end{equation} 

\subhead{Defense} To evaluate the performance of our proposed defenses, we calculate the Defense Rate (DR) for an attacked scene,  the Classification Drop (CD) for an unattacked scene, and the Post-Defense Accuracy (PDA) for any scene.

DR is defined as the probability that the defended image of a scene is classified as true class, given it is an attacked scene and its image was not classified as the true class before the defense,  
\begin{equation}
\text{DR}=Pr(C(D(I))=t^*|I=\alpha(S), C(I)\neq t^*)
\label{eq: DR}
\end{equation}

CD is defined as the probability that the image of an unattacked scene get misclassified after applying the defense. 
\begin{equation}
\text{CD}=\text{CA}-Pr(C(D(I))=t^*|I=S)
\label{eq: CD}
\end{equation}

PDA is defined  as the probability that  the image of any scene is correctly classified as the true class after the defense, 
\begin{align}
\text{PDA}&=(1-\pi_A)Pr(C(D(I))=t^*|I=S)\nonumber\\
&+\pi_APr(C(D(I))=t^*|I=\alpha(S))
\label{eq: PDA}
\end{align}

When U-ASR=1,  using equation \ref{eq: u-asr}, \ref{eq: DR} and \ref{eq: CD}, equation~\ref{eq: PDA} can be expressed as, 
\begin{equation} 
\text{PDA}=(1-\pi_A)(\text{CA}-\text{CD})+\pi_A \text{DR}
\label{eq: post-da}
\end{equation}

\section{Experimental Results}


To evaluate the performance of our proposed defenses, we conducted a series of experiments. The physical attack we attempt to defend against is the camouflage art attack proposed by Eykholt et. al~\cite{eykholt2018robust}.  The classifier we used to evaluate the the proposed defense was trained to differentiate 17 common US traffic signs using LISA traffic sign database~\cite{mogelmose2012vision} (a US traffic sign database). The classifier was reported to achieve 91\% classification accuracy in their paper.   We started by making a dataset composed of photos of unattacked ground truth source signs and physical attacked signs. Then we demonstrated the effectiveness of the proposed defense method under the three scenarios we discussed in Section~\ref{sec: scenarios}.  We assume $\pi_A=0.5$ in all scenarios. 

\subsection{Dataset}
To the best of  our knowledge, there exists no database that made  specifically for physical attack, especially using camouflage art attack.  A physical attack database should be constructed with the photos of the physically  attacked objects. This is because empirically we found that defenses against physical perturbations are very different from the digital simulation.  One reason is that the many effects introduced during capturing images of physically  attacked objects, such as   the curvature of surfaces, focus blur, sampling effects, sensor noise, will result in  significant discrepancies between physical perturbations  and digital approximation.  Therefore,  it is important to create a new database to fill this gap and benefit future research in the community. 
  
To make the database, we first purchased six US road signs which were included among the 16 which classes the LISA-CNN is trained to distinguish between.  These six signs are indicated above in  Table~\ref{tab: traffic signs} as `source' signs.  

To create training data for the attack, and assess the baseline performance of the LISA-CNN, we first captured a set of images of the six unattacked signs in our possession.  This was done by photographing each sign at angles running from $-50$ to 
$+50$ degrees in an increments of  $10$ degrees, to create a set of  $66$ images of unattacked signs.

Next, we launched a series of multi-sticker attacks against the six signs in our possession, using each of the 15 remaining classes listed in Table~\ref{tab: traffic signs} as the attack's target.  This was done by following the attack protocol described in~\cite{eykholt2018robust}.  For each pair of source and target signs, we first created a digital copy of the attacked sign.  This digital copy was projected onto the corresponding physical copy of the source sign, then black and white stickers were placed on the sign in regions indicated by the digitally attacked version.
Front facing images of all of the attacked signs were captured, then cropped to approximately  $340 \times 340$ pixels and saved as PNG files.  This resulted in a set of 90 images of physically attacked signs, each with a different source-target class  pairing. The database 
is publicly available at \url{https://drive.google.com/drive/folders/1qOmSubSOVY8JzB3KfXhDQ38ihoY5GExK?usp=sharing}.

\subsection{Baseline Evaluation of the Classifier and Attack}
To assess the baseline classification accuracy of the LISA-CNN classifier trained by Eykholt et al., we evaluated its performance on the unattacked signs captured as part of our database.  In this evaluation, the LISA-CNN achieved $100\%$ classification accuracy.  We note that Eykholt et al. reported a $91\%$ classification accuracy during their evaluation of this trained classifier.  In this paper, when reporting metrics that depend on classification accuracy, we use the value that we obtained since this classification accuracy is measured on the same set of road signs in the attack set. 
Furthermore, this corresponds to  more challenging test conditions for our defense, since perfect performance would need to bring the defense rate equal to this higher classification accuracy.

Next, we measured the baseline performance of the attack by using the LISA-CNN to classify the images of physically attacked signs in our database.  Our implementation of the camouflage art attack achieved a  $0.9556$ targeted attack success rate (T-ASR) and a $1.0000$ untargeted attack success rate (U-ASR).  This result verifies that we were able to reproduce the attack, and that this attack can successfully fool the classifier.  
\begin{table}[t]
	\centering
	\caption{ Source and Target Traffic signs. S denotes ``source" and T denotes ``target".}
	\resizebox{0.43\textwidth}{!}{
		\begin{tabular}{c c|c  c}
			\hline
			\textbf{Category}& \textbf{Sign Name} &\textbf{Category}& \textbf{Sign Name} \\\hline
			S \& T & crossing & T & added lane\\
			S \& T & stop &T& keep right\\
			S \& T & yield&		T &lane ends\\
			S \& T & signal ahead&	T & stop ahead\\ 
			S \& T &speed limit 25&	T& turn right\\
			S \& T & speed limit 45&T& school / limit 25\\
			T  & merge &	T& speed limit 30\\
			T& school	&	T& speed limit 35\\
			\hline
	\end{tabular}}
	\label{tab: traffic signs}
\end{table}
\begin{table}[t]
	\centering
	\caption{Non-blind evaluation of our proposed defenses.}
	\resizebox{0.4\textwidth}{!}{
		\begin{tabular}{lccc}
			\hline
			\textbf{Proposed defense} & \textbf{DR} &\textbf{CD} &\textbf{PDA} \\\hline
			RemapW  & 0.4339&0.0000 &0.7170\\
			RemapB & 0.4556 &0.0000 &0.7283\\
			RemapT  & \textbf{0.9222}&0.0000& \textbf{0.9611}\\
			Reconst & 0.6778 &0.0000 &0.8389\\\hline
	\end{tabular}}
	\label{tab: non-blind}
\end{table}

\subsection{Non-Blind} 
\label{subsec: non-blind}
In our first set of experiments, we evaluated our defenses' performance in the non-blind scenario.  We used the digital versions of the perturbation masks obtained while training the attack as the oracle defensive masks  known to the defender.  While these digital masks are not perfect ground truth locations of the actual perturbations they are sufficiently close to evaluate our experiment. 

Using these oracle masks, we evaluated the three perturbation remapping defenses remap to white (RemapB), black (RemapW), and threshold (RemapT) as well as the targeted region reconstruction (Reconst) defense. 
We note that the classification drop is always zero in this experiment because the defender always knows if an attack is present and can choose when not to apply the defense.

Table~\ref{tab: non-blind} shows the performance of our defenses in the non-blind scenario.  
Thresholded perturbation achieved strongest performance with the highest defense rate of 0.9222 and post-defense accuracy of 0.9611.   Since both the remap-to-white and remap-to-black strategies will only affect approximately half of the stickers added to an object, it is reasonable to expect that the thresholded perturbation remapping approach outperforms these approaches.  Reconstruction approach achieved second highest performance. We believe that lower defense rate is predominantly due to  the slight misalignment between the ideal digital perturbation masks and the true locations of the physical perturbations in the attacked images.

\begin{table}[t]
	\centering
	\caption{Evaluation of proposed defenses in semi-blind scenario}
	\resizebox{\textwidth}{!}{
		\begin{tabular}{lccc|lccc}
			\hline
			\textbf{Defense Strategies} & \textbf{DR} &\textbf{CD} &\textbf{PDA}&\textbf{Defense Strategies} & \textbf{DR} &\textbf{CD} &\textbf{PDA} \\\hline
			RemapW-Par(6) + MV&0.3989&0.3333&0.5328&RemapW-Par(6) + SF&0.4186&0.1667&0.6260 \\
			RemapB-Par(6) + MV&0.0794&0.1667&0.4563&RemapB-Par(6) + SF&0.0690&0.0000&0.5348\\ RemapT-Par(6) + MV&0.5174&0.3333&0.5921& RemapT-Par(6) + SF&0.6453&0.1667&0.7393\\\hline
			Reconst-Par(6) + MV&0.3560&0.0000&0.6780&	Reconst-Par(6) + SF&0.3514&0.0000&0.6757\\\hline

			Reconst-Seq-Rand(1) &0.2815 & 0.0556&0.6130& Reconst-Seq-Rand4) &0.6200&0.1112&0.7544\\
			Reconst-Seq-Rand(2) &0.4237&0.0556&0.6840& Reconst-Seq-Rand(5) &0.6648&0.1389&0.7630\\
			Reconst-Seq-Rand(3)&0.5350&0.0834&0.7250& Reconst-Seq(6) &0.7000&0.1667&0.7667\\\hline
			
			Reconst-Seq-Rank(1)&0.3780&0.0000&0.6890&Reconst-Seq-Gtd(1)& 0.6778 &0.0000 &0.8389\\
			Reconst-Seq-Rank(2)&0.6336&0.0000&0.8168&Reconst-Seq-Gtd(2) &0.6623&0.0333&0.8145\\
			 Reconst-Seq-Rank(3)&\textbf{0.7001}&\textbf{0.0000}&\textbf{0.8501}&Reconst-Seq-Gtd(3)&0.6855&0.0667&0.8094\\
			Reconst-Seq-Rank(4)&0.6667&0.0000&0.8333&Reconst-Seq-Gtd(4) &0.7022&0.1000&0.8011\\
			Reconst-Seq-Rank(5)&0.7000&0.0000&0.8500&Reconst-Seq-Gtd(5)&0.7044&0.1333&0.7856\\\hline
			
\textbf{Other Methods} & \textbf{DR} &\textbf{CD} &\textbf{PDA}&\textbf{Other Methods} & \textbf{DR} &\textbf{CD} &\textbf{PDA} \\\hline
			DW~\cite{hayes2018visible}&0.2222&0.0000&0.6111& Median Filter (kernel=7) ~\cite{xu2017feature}&0.3777&0.3333&0.5222\\
JPEG (QF=10)~\cite{dziugaite2016study} &0.1333&0.0000&0.5667&Local Smooth~\cite{naseer2019local}&0.0000&0.0000&0.5000\\\hline
	\end{tabular}}
	\label{tab: semi}
\end{table}

\subsection{Semi-blind}
\label{subsec: semi-blind}
To evaluate our defenses in the semi-blind scenario, we created a set of estimated defensive masks for each of the 15 possible target classes.  Each set of defensive masks contained six pairings of source and target sign, i.e. one for each source sign in our database that an attack could be launched against.

Next, we used these sets of defensive masks to evaluate our relevant defensive strategies.  The results of these experiments are shown in Table~\ref{tab: semi}.  We adopt the notation Par  and Seq to denote that a defense was applied either in parallel or sequentially, and ($k$) to denote the number of defensive masks used for defense. When defenses were applied in parallel, we use the notation MV to denote majority vote fusion and SF to denote softmax fusion. 
We use Rand and Rank to denote the random or ranked mask selection strategy.  
Additionally,  we use Gtd to denote a special ``Guaranteed scenario" 
in which the defensive mask with the correct source-target pair was always included and the remaining masks were randomly chosen.

Results in  Table~\ref{tab: semi} show that for any mask selection strategy, sequential reconstruction outperforms both parallel reconstruction and perturbation remapping. The defense using three defensive masks  selected using ranked activation (Reconst-Seq-Rank(3)) outperforms all other strategies and achieved the highest defense rate of 0.7001, highest post-defense accuracy of 0.8501, and  zero classification drop on  unattacked images. We note that Reconst-Seq-Rank(3) is statistically the same performance as Reconst-Rank(5), but it is more computationally efficient using less  masks.

\boldhead{Comparisons of  Defensive Mask Selection Strategies} For all values of  $k$,  the ranked selection strategy achieved a higher defense rate and post-defense accuracy 
than the random  selection strategy. This shows that using a well chosen subset of defensive masks  improves our system's performance.  Additionally, it reinforces our observation 
that important  information about the true source class and attack target class can be observed in the top few class activations.

To explore impacts from the correct source and target defensive mask, we ran another set of experiments for the ``Guaranteed scenario''. Compared to the Reconst-Seq-Rand($k$) strategy,  the Reconst-Seq-Gtd($k$) strategy always achieved higher defense rate,  post-defense accuracy, and lower classification drop for the same $k$. These results imply  that the inclusion of the estimated mask for the correct source-target class pair can significantly improve the performance of defenses.

Comparing the Ranked strategy with the ``Guaranteed scenario", Ranked results are in a higher post-defense accuracy for $k\geq2$.  The main reason is that Ranked produces a significantly lower classification drop.  These results suggest that using the ranked selection strategy not only can pick out the ``best''  subset of  defensive masks to use, but can also exclude those that deteriorate the classification accuracy of unattacked images.

This is reinforced by examining the classification drop as $k$ increases.  For both Reconst-Seq-Gtd($k$) and Reconst-Seq-Rand($k$), the classification drop increases as $k$ increases, thus hurting the overall post-defense accuracy.  This is likely because some masks that negatively effect the overall performance are included.  By contrast, Reconst-Seq-Rank($k$) does not suffer from the same decrease in classification drop because unlikely defensive masks that may hurt performance are excluded.

\boldhead{Comparisons with Related Defenses} 
We compared the performance of our proposed defenses with several existing defenses against physical domain attacks. These include distortions that are universally applied to an image such as JPEG compression and median filtering, as well as the more sophisticated digital watermarking (DW) defensive method and the local smooth approach.  While we evaluated the performance of the JPEG defense using multiple quality factors and the median filtering defense using multiple kernel sizes, we report  only the strongest results in the interest of space.

The results in Table~\ref{tab: semi} show that all of our proposed strategies with the reconstruction defense can significantly outperform each of these existing defenses.  The digital watermarking defense proved to be the strongest performing existing defense, with a defense rate of 0.2222 and a post-defense accuracy of 0.6111.  However, even when only one randomly chosen estimated defensive mask is used, our region reconstruction defense outperforms this approach. Our best performance achieved more than three times higher in defense rate and about 40\% more in post-defense accuracy than this approach.  The relatively poor performance of these existing defenses likely occurs because they are targeted to defend against the adversarial patch attack.  Since the multi-sticker camouflage art attack works in a different manner and exhibits different visual properties, these defenses are not as well suited to protect against this and similar attacks.


\subsection{Blind}
To evaluate our defenses in the blind scenario, we created  randomly chosen defensive masks using both the overlapping (OL) and non-overlapping (NOL) strategies.  

Table~\ref{tab: blind} shows the results of these experiments.  In each experiment, we identified the optimal window size and number of windows for use in these masks through a grid search.   We chose window size $w$ vary from 2, 4, 8 16 pixels. Next we  controlled the number of windows $m$ by randomly selecting a ratio of  total  number of  windows based on the given window sizes. 
 The results reported in Table~\ref{tab: blind} correspond to the pairing of $w$ and ratio that achieved the highest post-defense accuracy.  A detailed examination of the choice of $w$ and ratio  is provided later in this section.

From Table~\ref{tab: blind}, we can see the strongest performance in terms of all evaluation metrics was achieved using targeted region reconstruction applied in parallel using 100 random masks with non-overlapping windows in conjunction with majority vote decision fusion (NOL-Reconst-Par(100)~+~SF).    
Even though no information regarding the attack could be leveraged, this defense was still able to achieve  a defense rate of 0.4102 with a corresponding classification drop of 0.0017 and a post-defense accuracy of 0.7043.  Though performance is worse than in the semi-blind scenario, we are still able to outperform existing defenses in all evaluation metrics. 
 We note that the local region reconstruction defense uniformly outperformed the targeted perturbation remapping strategy, and for targeted region reconstruction, applying the defense in parallel outperformed sequential application of the defense.

Creating defensive masks using the non-overlapping strategy significantly improves our defense's performance over the overlapping approach (i.e. choosing window locations uniformly at random).  Furthermore, we note that performance increases as the number of randomly chosen masks increases.  
While this comes at the price of additional computation costs, in practice we found that  our proposed defense takes  0.4 seconds on average using 100 masks without any attempt at execution time optimization.

\boldhead{Effect of Size and Number of Windows}
To understand the effect that the window size and number of windows (or ratio) in each randomly chosen defensive mask has on our defense, 
we provide detailed results of our search over these parameters in Table~ \ref{tab: 5}. The symbol $\ast$ means when window size was 16  and ratio was 0.625, the  computed number of windows was not an integer. However,  it equals to when ratio was 0.5 if rounded down, and equals to when ratio was 0.75 when rounded up. 

The results show that the defense rate increases as the ratio (i.e  the number of windows) increases.  After a certain point, the classification drop also increases, resulting in a negative effect on the post-defense accuracy.  We also find that increasing the window size increases the defense rate up to a certain point, after which the defense rate begins to decrease. Additionally, after a certain point,  increasing the window size also leads to an increase in the classification drop and a decrease in the post-defense accuracy. 
In our experiments, we found that the optimal window size was 8 pixels and ratio was 0.625.  More importantly, when choosing the window size and the ratio (i.e  the number of windows),  the defender must balance the trade-off between interfering with the attack and interfering with the unattacked scene content used by the classifier.

\boldhead{Comparisons with Related Defenses} 
From Table~\ref{tab: blind}, we can see that applying region reconstruction in parallel using non-overlapping masks outperforms the existing defenses that were also considered in the semi-blind scenario 
(the performance of these defenses do not change in the blind scenario).  
This result holds true even when only six randomly generated non-overlapping  masks are used.  
\begin{table}[t]
	\centering
	\caption{Defense performance in the blind-scenario.}
	\resizebox{\textwidth}{!}{
		\begin{tabular}{lccc|lccc}
			\hline
			\textbf{Defense Strategies}  & \textbf{DR} &\textbf{CD} &\textbf{PDA} & \textbf{Defense Strategies} & \textbf{DR} &\textbf{CD} &\textbf{PDA}\\\hline
			OL-RemapT-Par(6) + MV & 0.1210 &0.2367&0.4422 & NOL-RemapT-Par(6) + MV&0.1236&0.1283&0.4976\\
			OL-RemapT-Par(6) + SF&0.1271&0.1650&0.4811& NOL-RemapT-Par(6) + SF&0.1255&0.1267&0.4994\\
			
			OL-RemapT-Par(100) + MV &0.0713&0.0333&0.5190&	NOL-RemapT-Par(100) + MV& 0.0482& 0.0500&0.4991\\
			OL-RemapT-Parallel(100) + SF& 0.0778& 0.0333&0.5222 & 	NOL-RemapT-Par(100) + SF& 0.0737& 0.0667&0.5035\\\hline
			
			OL-Reconst-Seq(6) & 0.2352& 0.1782&0.5286&	NOL-Reconst-Seq(6)& 0.1942& 0.0775&0.5584\\
			
			OL-Reconst-Seq(100) & 0.1588 & 0.8450&0.1569& 	NOL-Reconst-Seq(100)&0.2553&0.6300&0.3027\\

			OL-Reconst-Par(6) + MV &0.1836&0.0717&0.5560 & NOL-Reconst-Par(6) + MV&0.3444&0.0467&0.6488\\
			OL-Reconst-Par(6) + SF&0.1762&0.0350&0.5706& 	NOL-Reconst-Par(6) + SF&0.3501&0.0317&0.6593\\
			OL-Reconst-Par(100) + MV &0.1362&0.0000&0.5681&	NOL-Reconst-Par(100) + MV& 0.4129 & 0.0067& 0.7031\\
			OL-Reconst-Par(100) + SF&0.2415&0.0167&0.6124 &	NOL-Reconst-Par(100) + SF&\textbf{0.4102}&\textbf{0.0017}& \textbf{0.7043}\\
			\hline
			
			\textbf{Other Methods} & \textbf{DR} &\textbf{CD} &\textbf{PDA}&\textbf{Other Methods} & \textbf{DR} &\textbf{CD} &\textbf{PDA} \\\hline
			DW~\cite{hayes2018visible}&0.2222&0.0000&0.6111&Median Filter (kernel=7) ~\cite{xu2017feature}&0.3777&0.3333&0.5222\\
			JPEG (QF=10)~\cite{dziugaite2016study} &0.1333&0.0000&0.5667&Local Smooth~\cite{naseer2019local}&0.0000&0.0000&0.5000\\\hline
	\end{tabular}}
	\label{tab: blind}

	\centering
	\caption{Local region reconstruction using 100 parallel masks with  non-overlapping windows and softmax fusion. * For window size 16, ratio 0.625 results in a non-integer  number of windows. }
	\resizebox{\textwidth}{!}{
		\begin{tabular}{l| c |c| c| c}
			\hline
			&Ratio $=0.25$&Ratio $=0.5$&Ratio $=0.625$& Ratio $=0.75$\\
			& DR  \text{ } \text{ } \text{ } \text{ } CD  \text{ } \text{ } \text{ } 
			\text{ } PDA&DR \text{ }  \text{ } \text{ } \text{ } CD \text{ } \text{ } \text{ } \text{ }  PDA & DR \text{ } \text{ } \text{ } \text{ }  CD \text{ } \text{ } \text{ } \text{ } PDA&DR \text{ } \text{ } \text{ } \text{ } CD \text{ } \text{ } \text{ } \text{ } PDA\\\hline
			$w=2$&0.0546 \text{ } 0.0000 \text{ } 0.5273 & 0.1528 \text{ } 0.0000 \text{ } 0.5764& 0.2027 \text{ } 0.0033 \text{ } 0.5997& 0.2770 \text{ } 0.1283 \text{ } 0.5744\\
			$w=4$& 0.0537 \text{ } 0.0000 \text{ } 0.5269 & 0.1818 \text{ } 0.0000 \text{ } 0.5909 & 0.2395 \text{ } 0.0000 \text{ } 0.6198 &0.3153 \text{ } 0.0233 \text{ } 0.6460\\
			$w=8$& 0.1648 \text{ } 0.0000 \text{ } 0.5824 &0.3268 \text{ } 0.0000 \text{ } 0.6634&\textbf{0.4102 \text{ } 0.0017 \text{ } 0.7043}& 0.4701 \text{ } 0.2500 \text{ } 0.6101\\
			$w=16$&0.0183 \text{ } 0.0000 \text{ } 0.5092 &0.1073 \text{ } 0.1400 \text{ } 0.4837& $\ast$ & 0.1353 \text{ } 0.3333 \text{ } 0.401\\\hline
	\end{tabular}}
	\label{tab: 5}
\end{table}

\section{Conclusions}
In this paper, we proposed new defense strategies against physical domain attacks with a special focus on the multi-sticker attacks, like camouflage art attack. Our proposed methods attempt to maximize the likelihood of diminishing the effect of the physical perturbation, given the defender's different levels of knowledge. We conducted an extensive amount of experiments to show that our proposed defense can successfully defend against  the camouflage art attack under many scenarios with small classification drops on unattacked objects. Additionally,  we built a new database using the camouflage art attack that contains photos of 90 physical attacked traffic signs and six source signs. This database may benefit future research in the community.

%
%
\bibliographystyle{splncs04}
\bibliography{citation/citations.bib}
\end{document}